\documentclass[iop,apj,tighten,numberedappendix]{emulateapj}
\usepackage[breaklinks,colorlinks,urlcolor=blue,citecolor=blue,linkcolor=blue]{hyperref}
\usepackage{amsmath,amssymb}
\usepackage{color}

\def\comment#1{}

\usepackage{ulem} 

\begin{document}

\title{Indication for a Compact Object Next to a LIGO--Virgo Binary Black Hole Merger}

\author
{Shu-Cheng Yang\altaffilmark{1}, 
Wen-Biao Han\altaffilmark{1,2,3,4,5}, 
Hiromichi Tagawa\altaffilmark{1}, 
Song Li\altaffilmark{1,3}, 
Chen Zhang\altaffilmark{6}, 
}
\affiliation{\altaffilmark{1}Shanghai Astronomical Observatory, Chinese Academy of Sciences, Shanghai 200030, People's Republic of China;wbhan@shao.ac.cn\\
\altaffilmark{2}Hangzhou Institute for Advanced Study, University of Chinese Academy of Sciences, Hangzhou 310124, People's Republic of China\\
\altaffilmark{3}School of Astronomy and Space Science, University of Chinese Academy of Sciences, 
Beijing 100049, People's Republic of China\\
\altaffilmark{4}Taiji Laboratory for Gravitational Wave Universe (Beijing/Hangzhou), University of Chinese Academy of Sciences, Beijing 100049, People's Republic of China\\
\altaffilmark{5}Key Laboratory of Radio Astronomy and Technology, Chinese Academy of Sciences, A20 Datun Road, Chaoyang District, Beijing 100101, People's Republic of China\\
\altaffilmark{6}Shanghai University of Engineering Science, Shanghai, 201620, People's Republic of China\\
}



\begin{abstract}
The astrophysical origin of binary black hole (BBH) mergers remains uncertain, although many events have been observed by the LIGO–Virgo–KAGRA network. Such mergers are potentially originated in the vicinity of massive black holes (MBHs). GW190814, due to its secondary mass and mass ratio being beyond the expectations of isolated stellar evolution theories, is a promising event that has occurred in an active galactic nucleus (AGN) disk. In this model, a compact object resides in the vicinity of a merging BBH. Here we report multiple pieces of evidence suggesting that GW190814 is a BBH merging near a compact object. The orbital motion of BBHs around a third body produces a line-of-sight acceleration (LSA) and induces a varying Doppler shift. Using a waveform template that considers LSA, we perform Bayesian inference on a few BBH events with a high signal-to-noise ratio in the gravitational-wave (GW) transient catalog. Compared to the model for isolated BBH mergers, we obtain significantly higher network signal-to-noise ratios for GW190814 with the inclusion of  LSA, constraining the LSA to $a = 0.0015^{+0.0008}_{-0.0008} ~c~\mathrm{s}^{-1}$ at a $90 \%$ confidence level. Additionally, the Bayes factor  for the LSA case over the isolated case is $58/1$, indicating that the GW data strongly prefer the LSA model. We conclude that this is the first indication showing merging BBHs are located near a third compact object.
\end{abstract}

\section{Introduction}

The detection of gravitational waves (GWs) from merging binary black holes (BBHs) and neutron stars has opened a new era of astronomical and physical research \citep{ligo2016observation, ligo2017gw170817}. So far, the ground-based detectors, i.e., the Advanced LIGO \citep{ligo2015advanced}, Advanced Virgo \citep{acernese2014advanced}, and KAGRA \citep{2021PTEP.2021eA101A}, have reported more than 90 GW events with high signal-to-noise ratios, most of which are BBHs\citep{ligo2019gwtc1, ligo2021gwtc2, ligo2021gwtc3}. 
Next-generation detectors such as the Einstein Telescope
\citep{punturo2010einstein} and Cosmic Explorer \citep{reitze2019cosmic} will further improve the detector sensitivity and are expected to detect millions of BBHs every
year\citep{reitze2019cosmic}, out to a redshift as high as $\sim 20$ \citep{pieroni2022detectability}. In about a decade, space-borne detectors such as the Laser Interferometer Space Antenna \citep{amaro2017laser}, Taiji \citep{hu2017taiji}, and TianQin \citep{luo2016tianqin} will begin to observe low-frequency signals. They are capable of detecting BBHs in their early evolutionary stages when the binaries are emitting milli-Hertz GWs \citep{sesana2016prospects}. Such observations will provide us with a more complete view of the formation and evolution of BBHs.

\begin{figure}
\centering
\includegraphics[width=85mm]
{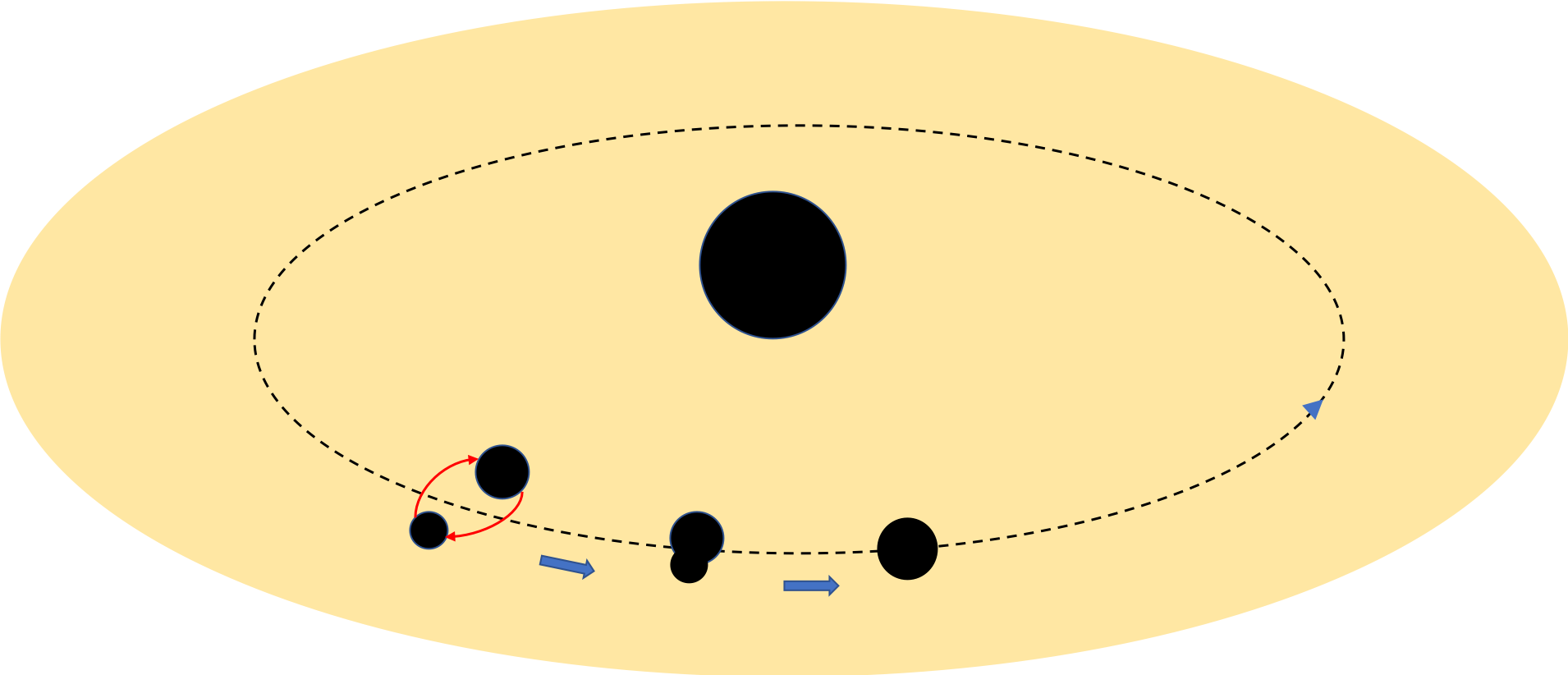}
\caption{BBH forms around a compact object and later coalesces.}
\label{fig:diagram}
\end{figure}

Theoretically, BBHs could form due to either binary star evolution or stellar dynamical interactions in star clusters \citep{abbott2016astrophysical, abbott2019binary, abbott2020properties}. Besides these two conventional formation channels, a third scenario, mergers in active galactic nucleus(AGN) disks \citep{mckernan_ford_2012, peng21}, has recently gained much attention. In addition, stellar-mass objects can be tightly bound to merging BBHs in these environments as explained below. As illustrated in Figure~\ref{fig:diagram}, this scenario suggests that BBHs may form and coalesce in the vicinity of a compact object (possibly a stellar-mass black hole, BH). A large fraction of the BBHs produced in this third channel could grow to be as massive as $\sim 30-100~M_\odot$ (see ~\onlinecite{arca-sedda23nuclei} for a review), which is consistent with LIGO/Virgo/KAGRA observations. The estimated event rate is also compatible with the current detection rate of BBHs.

Unlike the BBHs forming in isolation, BBHs close to a third compact object are moving within a deep gravitational potential and are therefore accelerating relative to a distant observer. Such a ''peculiar acceleration'' could induce several observable signatures in the GW signal. First, it changes the line-of-sight velocity of the GW source and hence modulates the GW frequency due to the Doppler effect \citep{YunesEtAl2011,bonvin17,inayoshi_tamanini_2017,meiron_kocsis_2017}. Second, the acceleration also induces a shift in the GW phase due to an aberrational effect \citep{torres-orjuela_chen_2020a, bonvin2022aberration}. Third, the amplitude of the GW also varies since the GW radiation is beamed in the direction of the orbital motion \citep{torres2019detecting, torres2023moving}. From an astrophysical perspective,  BBHs could form within a distance of $10$ Schwarzschild radii from a massive black hole(MBH) \citep{chen_han_2018,addison_gracia-linares_2019,peng21}, or BBHs could merge during very hard binary-single interactions \citep{samsing2022agn,Tagawa2021_ecc}. Nevertheless, a recent search using the neutron-star mergers detected by LIGO-Virgo-KAGRA did not find any significant acceleration, which places an upper limit of about ${\cal O}(1)\,{\rm km\,s^{-2}}$ to the line-of-sight acceleration (LSA) of the source \citep{vijaykumar23}. Such a limit indicates that neutron-star binaries should reside at distances greater than 
$50(10^6\, M_\odot/M_{\mathrm{MBH}})$ Schwarzschild radii from an MBH, where $M_{\mathrm{MBH}}$ is the mass of the MBH. 

Due to the Doppler effect, an accelerating source causes the detector frame mass to change over time, leaving an imprint on the GW waveform \citep{vijaykumar23, wong2019binary, randall2019direct, tamanini2020peculiar}. 
By adopting the phase correction of GW waveforms given by ~\onlinecite{bonvin17}, we search for a possible third body close to the LIGO-Virgo-KAGRA BBHs. In particular, we apply Bayes inference on the LIGO-Virgo-KAGRA data and use the resulting
Bayes factors to quantify the significance of a possible LSA. We choose several GW events with a high network signal-to-noise ratio($>20$) from the GWTC. We find that the Bayes factor $ \mathrm{BF}^{\mathrm{LSA}}_{\mathrm{iso}}$, which indicates whether the waveform model with an LSA is preferred relative to the model for an isolated BBH, is not large for the majority of the events. This indicates that, for most of the events, there is no clear preference between the two models. On the other hand, for GW190814, we find $\mathrm{BF}^{\mathrm{LSA}}_{\mathrm{iso}} = 58 / 1$, which suggests that the waveform model with an LSA is strongly \citep{kass1995bayes} preferred by the data from GW190814.

\section{Bayesian inference with the LSA}

\subsection{Waveform Template}
A binary with a total mass $M_{\rm src}$ in the source frame, when considering cosmological redshift, Doppler shift, and LSA, the total mass in the detector frame given by \citep{vijaykumar23}
\begin{equation}
M_{\mathrm{det}} = M_{\mathrm{src}}(1 + z_{\mathrm{cos}}) (1 + z_{\mathrm{dop}}) (1 + a/c \times t),
\end{equation}
where  $z_{\mathrm{cos}}$ is the cosmological redshift of the source and $z_{\mathrm{dop}} \sim v/c$ is the Doppler shift due to a constant line-of-sight velocity $v$ of the source. The equation above assumes that $z_{\mathrm{dop}} \ll 1$ and $|a/c|\times t \ll 1$. Therefore, an accelerating source produces a time-varying detector-frame mass, which leaves an imprint on the gravitational waveform. In this work, we use a waveform template that includes the LSA. 

The waveform model is based on IMRPhenomXPHM \citep{pratten2021computationally}, which includes higher modes ($(l, |m|) = (2,1), (3,3),(3,2), (4,4)$) in addition to the dominant harmonic modes (2,2) in the precessing frame. In particular, our models are constructed by adding a modification term to Equation~(2.3) in ~\onlinecite{santamaria2010matching}. For comparison, we also show the network signal-to-noise results from the waveform model IMRPhenomPv2 \citep{hannam2014simple, husa2016frequency}, which does not consider higher-order multipoles(see Figure~\ref{fig:diagram3}).

When considering the LSA effect, since the (2,2) mode is dominant, we now focus only on the LSA modification for this mode, and the phase modification is\citep{vijaykumar23}
\begin{equation}
\Delta \Psi(f)_{22} = \frac{25}{65536 \eta^{2}}\left(\frac{GM}{c^3}\right)\left(\frac{a}{c}\right){\nu_{f}}^{-13},
\label{eqn:delta_psi}
\end{equation}
where $M$ is the total mass of the binary with masses $m_1$ and $m_2$, $\nu_{f} = (\pi GMf/c^{3})^{1/3}$, and $\eta = m_1 m_2 / M^2$ is the symmetric mass ratio. Considering the signal-to-noise ratio of other modes in GW190814 is much less than that of the $m = 2$ modes, the influence of LSA for higher modes on the PE could be ignored, and it would not change our conclusions.

\subsection{Parameter Estimation and Model Selection}
In the parameter estimation of GW signals of BBH, Bayesian analysis is widely used \citep{veitch2015parameter, puecher2022testing}. Consider detector data $d$ and a hypothesis $\mathcal{H}$, and prior probability distribution $ p(\boldsymbol{\theta}|\mathcal{H})$. From Bayes's theorem, the posterior distribution $p(\boldsymbol{\theta}|d, \mathcal{H})$ has the form \citep{veitch2015parameter}
\begin{equation}
p(\boldsymbol{\theta}|d, \mathcal{H}) = \frac{p(d|\boldsymbol{\theta}, \mathcal{H}) p(\boldsymbol{\theta}|\mathcal{H})}{p(d|\mathcal{H})}.
\end{equation}
Typically, there are many parameters in the models as $\boldsymbol{\theta} = \{\theta_1, \theta_2... \theta_N\}$. The joint probability distribution in the multi-dimensional space $p(\boldsymbol{\theta}|d, \mathcal{H})$ describes the collective knowledge about all parameters and their relationships. For a specific parameter, we can obtain its result by marginalizing over the other unwanted parameters: 
\begin{equation}
p(\theta_1|d, \mathcal{H}) = \int p(\boldsymbol{\theta}|d,\mathcal{H}) \mathrm{d}\theta_2...\mathrm{d}\theta_{N}.
\end{equation}
The term $p(d|\boldsymbol{\theta}, \mathcal{H})$ is the likelihood, and in our background, it takes the form
\begin{equation}
p(d|\boldsymbol{\theta}, \mathcal{H}) \propto \exp{\left[-\frac{1}{2} <d - h(\boldsymbol{\theta})|d - h(\boldsymbol{\theta})>\right]},
\end{equation}
where the noise-weighted inner product $<\cdot|\cdot>$ can be written as
\begin{equation}
< a|b > = 4 \mathrm{Re} \int_{f_{\mathrm{max}}}^{f_{\mathrm{min}}} \frac{\tilde{a}^{\star}(f)\tilde{b}(f)}{S_n(f)}\mathrm{d}f,
\end{equation}
where $\tilde{a}(f)$ is the Fourier transform of $a(t)$, $\tilde{a}^{*}(f)$ is the complex conjugate of $\tilde{a}(f)$, and $S_{n}(f)$ is the power spectral density of the GW detectors' noise, which was obtained from data prior to the GW events.
The evidence $Z$ is the fully marginalized likelihood multiplied by the prior over all parameters of the model $\mathcal{H}$,
\begin{equation}
Z = p(d|\mathcal{H}) = \int p(d|\boldsymbol{\theta}, \mathcal{H}) p(\boldsymbol{\theta}|H) \mathrm{d}\theta_1...\mathrm{d}\theta_{N}.
\end{equation}
Based on the available data, the Bayes factor is a measure of the evidence in favor of one hypothesis over another. It is the ratio of the evidence for the data given one hypothesis to 
that given the other hypothesis. The Bayes factor is used to compare the strength of evidence between two competing models or hypotheses. A larger Bayes factor indicates stronger evidence favoring one hypothesis over the other. For example, we can compare the evidence for Model A and Model B by calculating the Bayes factor\citep{thrane2019introduction}
\begin{equation}
\mathrm{BF}^{\mathrm{A}}_{\mathrm{B}} = \frac{\mathcal{Z}_{\mathrm{A}}}{\mathcal{Z}_{\mathrm{B}}}.  
\end{equation}
To quantify the evidence for the presence of a third body in GW events, we calculate the ${\rm LSA} / {\rm iso}$ Bayes factors, i.e. $\mathrm{BF}^{\mathrm{LSA}}_{\mathrm{iso}}$. The result $\mathrm{BF}^{\mathrm{LSA}}_{\mathrm{iso}} > 10 / 1$ indicates that the GW data strongly prefer the LSA template \citep{kass1995bayes}. In this work, to perform parameter estimation and model selection, we used BILBY\citep{ashton2019bilby}, a Bayesian inference software designed for GW astronomy.

\section{Results}

\begin{figure*}
\centering
\includegraphics[width=160mm]
{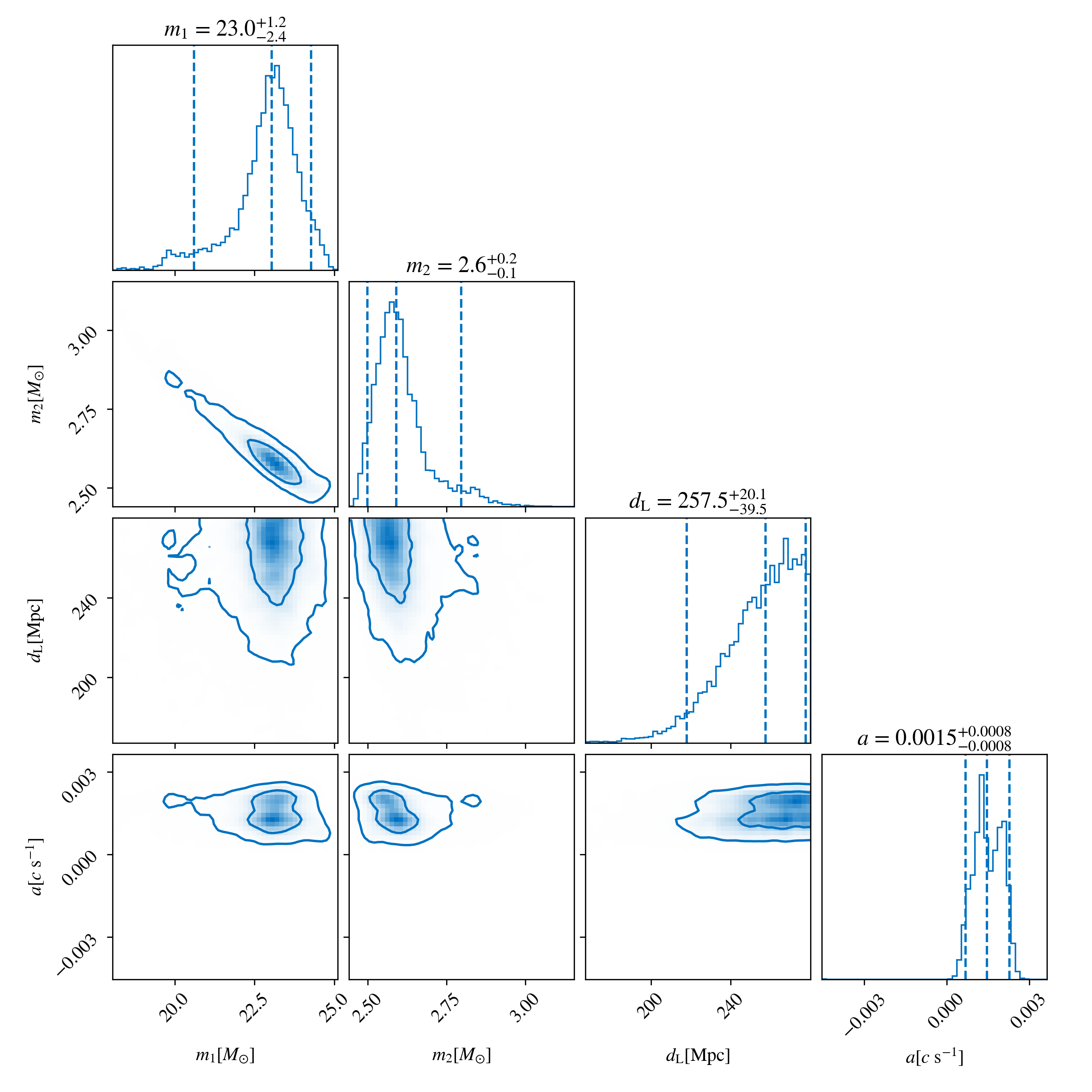}
\caption{Posterior distributions for the primary and secondary masses, luminosity distance, LSA for GW190814. The contours show the 50\% and 90\% credible intervals, and the vertical dashed lines mark the median and 90\% credible intervals. Notice that the BH masses are given in the rest frame of the source.  The waveform template is based on IMRPhenomXPHM.}
\label{fig:diagram2}
\end{figure*}

\begin{figure}
\centering
\includegraphics[width=85mm]
{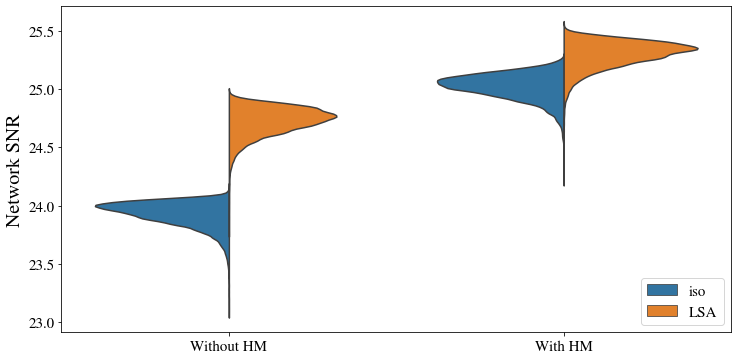}
\caption{Posterior distributions for network signal-to-noise ratio of GW190814 for the isolated BBH model and the LSA model. The posterior distributions for the isolated BBH and the LSA models are shown in blue and orange, respectively. "Without HM" or "With HM" means if the waveform includes the higher multipole.
}
\label{fig:diagram3}
\end{figure}

We perform a full parameter estimation on a waveform template that includes LSA. Higher-order modes play a crucial role in the parameter estimation of asymmetric systems like GW190814 \citep{varma2014gravitational}. Therefore, our waveform is based on IMRPhenomXPHM, which considers higher-order modes. In addition, in this waveform,  the spin and spin precession of the binary are considered, while the eccentricity is not considered. Figure~\ref{fig:diagram2} shows the estimation results of some parameters for GW190814(for all parameters and their prior settings, please refer to the Appendix.). The median value of LSA is estimated to be about $0.0015 ~c~\mathrm{s}^{-1}$. This non-zero LSA produces a variation in redshift on the waveforms, which can then be identified through the data analysis. In addition, GW190814 is thought to have emitted higher-multipole radiation \citep{gw190814}. To ensure the validity of our analysis, we use the results that do not consider higher multipoles as a comparison. In Figure~\ref{fig:diagram3}, the label "Without HM" or "With HM" indicates whether the waveform includes the higher multipole. As shown in Figure~\ref{fig:diagram3}, for both cases, the network signal-to-noise ratios obtained using the LSA BBH model are higher than those obtained for the isolated BBH model. Furthermore, as expected, the models considering higher multipoles also have higher signal-to-noise ratios. 

Environmental effects are also modulators of binary evolution, primarily quantified through frequency-dependent dephasing between vacuum and environment-modified waveforms. These phase discrepancies, typically parameterized via power-law dependencies analogous to post-Newtonian corrections, arise from environment-mediated alterations to orbital frequency evolution and phase accumulation\citep{Cutler_94, Blanchet_14}. Recent theoretical advances demonstrate that such dephasing signatures hold dual astrophysical significance: as discriminants between binary formation mechanisms and as probes of matter dynamics in extreme spacetime curvature\citep{Basu_24, Caputo_20, Caneva_24}.

For stellar mass GW sources in the dynamical and AGN channel, there will be dephasing in GW signals because of different environmental effects. Based on \citep{zwick2025environmental}, Roemer delays arise from Doppler acceleration in triple systems due to gravitational acceleration from a tertiary body, which is the scenario mainly discussed in this paper. Tidal binding energy effects modify the binary's orbital energy via tidal forces from a nearby massive body, affecting the inspiral timescale. Bondi-Hoyle-Lyttleton(BHL) drag in the subsonic limit and dynamical friction drag in the supersonic limit (supersonic drag) describe gas-induced frictional forces on binaries embedded in AGN discs. Circumbinary disc torques result from viscous interactions between the binary and a surrounding accretion disc, driving orbital decay and phase deviations. These mechanisms are crucial for understanding environmental impacts on GW signals and have different frequency power laws\citep{cole2022disks}. Therefore, the exponent of $v_f$ in Equation~(\ref{eqn:delta_psi}) could be used to distinguish environmental effects on BBH gravitational waveforms. Therefore, we rewrite Equation~(\ref{eqn:delta_psi}) in the following form,

\begin{equation}
\Delta \Psi(f)_{22} = A\nu_{f}^{-n_I}.
\end{equation}
Therefore, in this formula, different values of $n_I$ represent different environmental effects for BBHs. 
Tidal binding energy has $n_I=11$, BHL drag $n_I=14$, and both supersonic drag and circumbinary disc torques feature $n_I=16$. In the case of Roemer delays, $n_I = 13$, and the coefficient $A$ now is related to the line of sight acceleration, i.e.
\begin{equation}
 A = \frac{25}{65536 \eta^{2}}\left(\frac{GM}{c^3}\right)\left(\frac{a}{c}\right).
\end{equation}

\begin{figure*}
\centering
\includegraphics[width=160mm]
{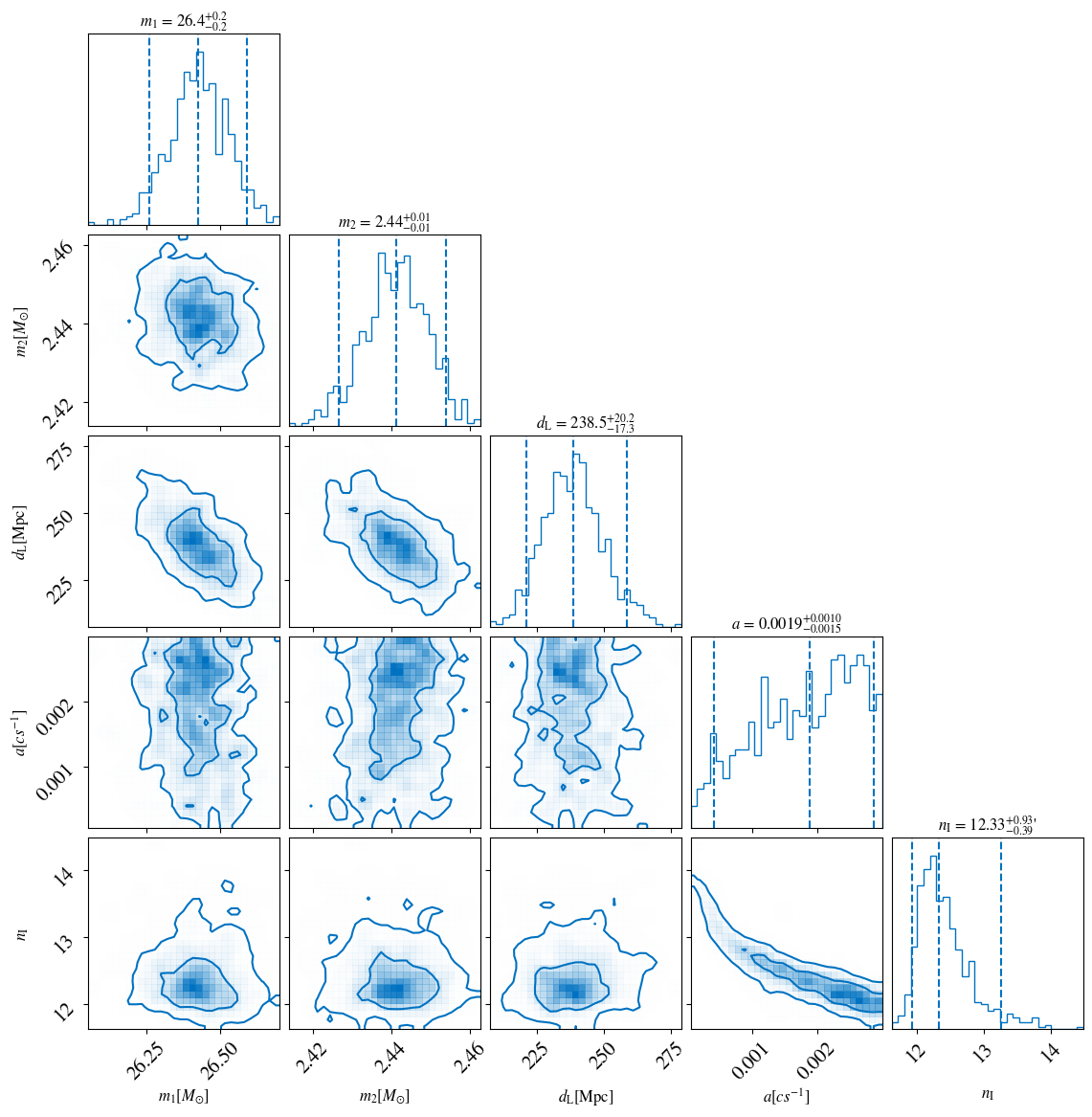}
\caption{Posterior distributions for the primary and secondary masses, luminosity distance, LSA and $n_I$ for GW190814. The contours show the 50\% and 90\% credible intervals, and the vertical dashed lines mark the median and 90\% credible intervals. Notice that the BH masses are given in the rest frame of the source.  The waveform template is based on IMRPhenomXPHM.}
\label{fig:diagram5}
\end{figure*}

We perform a parameter estimation with $n_I$, and the results are shown in the 
Figure~\ref{fig:diagram5}. From this figure, the posterior peaks at $n_I$ = 12.3 and 13 are indeed within the $90\%$ confidence interval, which indeed seems to favor our LSA hypothesis. Our result also tends to excludes the effects from other astrophysical environmental factors and the first eccentricity correction to the GW phase scales as $\sim f^{-11/3}$ mentioned by \cite{zwick2025environmental}.

All of these results suggest that introducing a nonzero LSA (or equivalently, a third compact object) may be reasonable. 

The LSA we estimated is positive, indicating that the direction of the LSA points away from the observer. In the positive (negative) LSA case, when the binary is moving away from (toward) the observer, i.e. in the redshift (blueshift) case, its line-of-sight velocity will increase (decrease), and in both cases, the chirp mass in the detector frame will be heavier during the merger. To be more quantitative, for a signal duration of about 10 seconds, the difference in the line-of-sight velocity is about  $|\Delta v| \sim 0.0015~c~\mathrm{s}^{-1} \cdot 10~s =0.015~c$. The corresponding difference in the redshift is about $\Delta z \sim \Delta v/c/2 \sim 0.008$. Therefore, we expect a difference of $M_{\rm chirp} \cdot \Delta z \sim 0.05~\mathrm{M_{\odot}}$ in the measured chirp masses, which is likely within the estimation error and may be hard to observe with the current GW detectors.

\section{Conclusion and Discussion}

Based on the results above, we assume that the event GW190814 is likely merging very close to a compact object. The orbital motion induces a varying redshift and produces detectable dephasing in GWs. The strong evidence of LSA for this event supports this assumption. Compared to other BBH events, GW190814 \citep{gw190814} has a larger asymmetric mass ratio and longer inspiral signals in the LIGO band \citep{gw190814}, which aids in the detection of the LSA through GWs.

From an astrophysical standpoint, the conditions and mechanisms by which a binary system approaches such proximity to an MBH before merging raise intriguing questions. The environment around supermassive BHs is often complex, with accretion disks, dense stellar populations, and strong gravitational perturbations. The merger's occurrence within this setting provides a rare observational glimpse into the interplay of these factors. 

We propose another possibility that the third body could be a stellar-mass BH (rather than an MBH, as proposed by \citealt{peng21}) to explain the acceleration of $a\sim 0.001~c~{\rm s}^{-1}$. For a BBH to experience this acceleration, the third body needs to reside at $\sim 1500~r_{\rm sch} (m_3/20~M_{\odot})^{-1/2}(a/0.001~c~{\rm s}^{-1})^{-1/2}$ from the merging BBH, where $m_3$ is the mass and $r_{\rm sch}$ is the Schwarzschild radius of the third body. Additionally, binaries can be hardened down to $s_{\rm min}\sim 10^{10}~{\rm cm}$ $(v_{\rm disp}/600~{\rm km/s})^{-2}$ through binary-single interactions, which is limited by the ejection from host systems due to kicks during these interactions (e.g. \onlinecite{samsing2017EccentricBH}), where $v_{\rm disp}$ is the velocity dispersion of the host system, and $1500~r_{\rm sch} \sim 10^{10}~{\rm cm}(m_3/20~M_{\odot})$. 
Furthermore, if the merger occurs via the GW capture mechanism during binary-single interactions, the third body can be bound to the merging BBH at a distance of $\gtrsim s_{\rm min}$ from the center of the merging BBH \citep{samsing2017EccentricBH}. Therefore, if a merger occurs through the GW capture mechanism in a system with a deep gravitational potential of $v_{\rm disp}\gtrsim 600~{\rm km/s}$ as predicted for the AGN channel \citep{samsing2022agn}, the acceleration of $\sim 0.001~c~{\rm s}^{-1}$ becomes possible. 

In fact, $\sim 2\%$ and $\sim 30\%$ of all mergers and the GW capture mergers following binary-single interactions result in the accelerations of $\geq 0.001~c~{\rm s}^{-1}$ in the fiducial model of \citet{Tagawa2021_ecc}. Hence, the third body can be a stellar-mass BH to explain the observed acceleration. Whether the third body is an MBH or a stellar-mass BH, we consider that the AGN channel to be a promising model for explaining the significant acceleration as no other models have been proposed that effectively account for the high acceleration and the high event rate of GW190814 ($\sim 1$--$23~{\rm Gpc}^{-3}{\rm yr}^{-1}$, \citealt{gw190814}). 

In this Letter, we present evidence for a BBH merger occurring near a compact object. However, simulations and theoretical studies focusing on the dynamics of binary systems in the vicinity of BHs will be beneficial. A more precise waveform template and enhanced data analysis will improve parameter estimation. We believe that during the O4 run and future observations, more such events will be discovered. Collaborations with electromagnetic observations, both ground and space-based, can assist in identifying any counterpart signals (e.g., gamma-ray bursts, X-ray emissions, optical flares) that may be associated with these unique events \citep{Tagawa2023}. This multimessenger approach would further unveil the environments of GW events and enrich our understanding of the underlying physical processes.

\begin{acknowledgments}
This work is supported by the National Key R\&D Program of China (grant No. 2021YFC2203002, 2024YFC2207700), NSFC (National Natural Science Foundation of China) No. 12473075, 11773059, and 12173071. This work made use of the High Performance Computing Resource in the Core Facility for Advanced Research Computing at Shanghai Astronomical Observatory. We thank Qian Hu at the University of Glasgow for the help with the Bayes inference. We also thank Xian Chen's help in the astrophysical background.
\end{acknowledgments}

%






\appendix



In this work, there are 16 parameters in total\citep{ashton2019bilby}, among which 8 are intrinsic (the two BH masses $m_{1,2}$, their dimensionless spin magnitudes $a_{1,2}$, the tilt angle between their spins and the orbital angular momentum $\theta_{1,2}$, the two spin vectors describing the azimuthal angle separating the spin vectors $\Delta \phi$, and the cone of precession about the system's angular momentum $\phi_{JL}$) describing the GW source system itself, and 8 are extrinsic (the luminosity distance $d_{\rm L}$, the R.A.,  the Decl., the inclination angle between the observer line of sight and the orbital angular momentum $\iota$, the polarization angle $\psi$, the phase at coalescence $\phi$, the time of coalescence $t_c$, and the LSA $a_{\rm L}$) describing the parameters related to GW propagation.

\begin{figure*}
\centering
\includegraphics[width=180mm]
{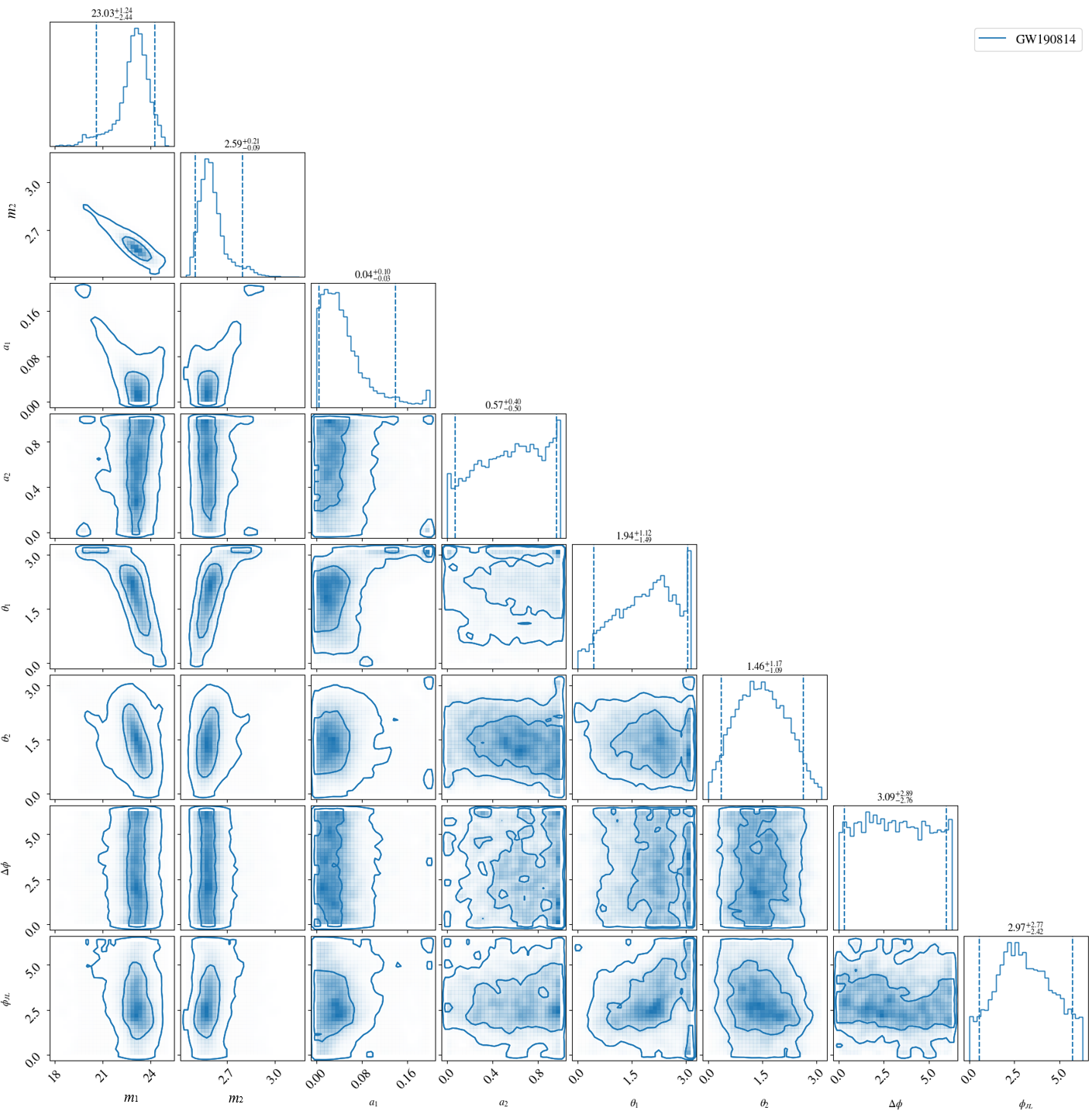}
\caption{Posterior distributions for intrinsic parameters (the two BH masses $m_{1,2}$, their dimensionless spin magnitudes $a_{1,2}$, the tilt angle between their spins and the orbital angular momentum $\theta_{1,2}$, the two spin vectors describing the azimuthal angle separating the spin vectors $\Delta \phi$, the cone of precession about the system's angular momentum $\phi_{JL}$) for GW190814. The contours show the $50\%$ and $90\%$ credible intervals, and the vertical dashed lines mark the 90\% credible intervals. Notice that the BH masses are given in the rest frame of the source. The waveform template is based on IMRPhenomXPHM.}
\label{fig:GW190814_All1}
\end{figure*}

\begin{figure*}
\centering
\includegraphics[width=180mm]
{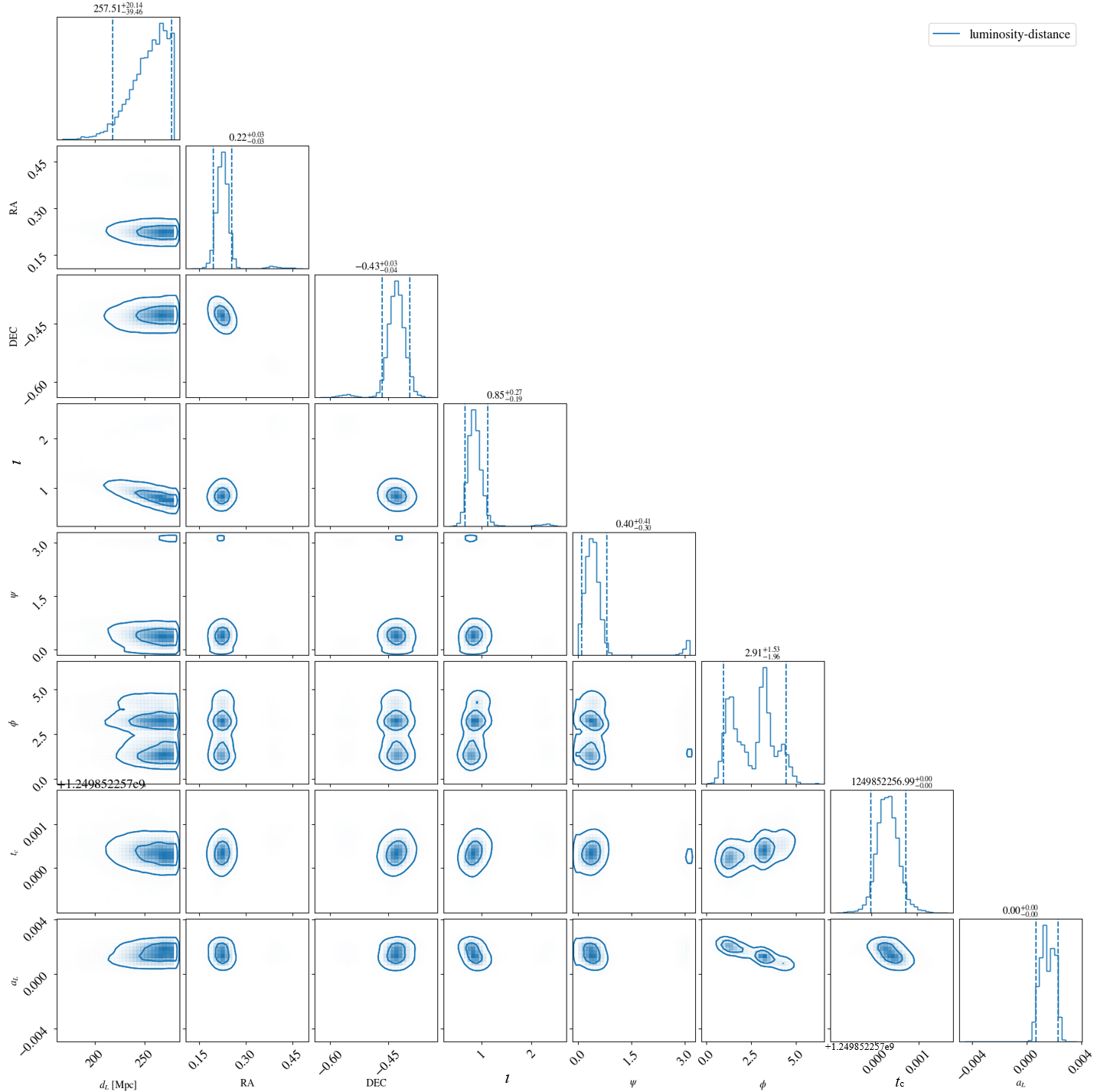}
\caption{Posterior distributions for extrinsic parameters (the luminosity distance $d_{\rm L}$, the R.A.,  the Decl., the inclination angle between the observer line of sight and the orbital angular momentum $\iota$, the polarization angle $\psi$, the phase at coalescence $\phi$, the time of coalescence $t_c$, and the line-of-sight acceleration $a_{\rm L}$) for GW190814. The contours show the 50\% and 90\% credible intervals, and the vertical dashed lines mark the $90\%$ credible intervals. The waveform template is based on IMRPhenomXPHM.}
\label{fig:GW190814_All2}
\end{figure*}

For GW190814, we perform a full parameter estimation on a waveform template that includes LSA.  Figure~\ref{fig:GW190814_All1} shows the estimation results of intrinsic parameters, and Figure~\ref{fig:GW190814_All2} shows the estimation results of extrinsic parameters. The setting of prior values can be found in Table~\ref{tab:table1}.

\begin{table*}
\begin{center}
\caption{The setting of prior values}
{\begin{tabular}{ccccc} \toprule
parameter & lower limit & upper limit & unit & distribution type\\
\hline
\rm{chirp~mass}        & $6$  &$7$ & $M_{\odot}$ & Uniform \\
\rm{mass~ratio}        & $0.1$  &$1.0$ & 1 & Uniform\\
$m_{1}$           & $15$  &$35$ & $M_{\odot}$ &  Constraint\\
$m_{2}$           & $1$  &$5$ & $M_{\odot}$ &  Constraint\\
$a_{1}$           & $0$  &$0.2$ & 1 & Uniform\\
$a_{2}$           & $0$  &$1.0$ & 1 & Uniform\\
$\theta_{1}$     & $0$  &$\pi $  & 1  & Sine\\
$\theta_{2}$      & $0$  &$\pi $  & 1  & Sine\\
$\Delta \phi$    & $0$  &$2 \pi $  & 1  & Uniform\\
$\phi_{\rm{JL}}$     & $0$  &$2 \pi $  & 1  & Uniform\\
$\rm{d_L}$        & $167$  &$280$ & $\rm{Mpc}$ &  PowerLaw($\alpha = 2$)\\
$\rm{Decl.}$             &  $ -\pi $  & $ \pi $  & 1  & Cosine\\
$\rm{R.A.}$              & $0$  &$2 \pi $  & 1 & Uniform \\
$\iota$       & $0$  &$ \pi $  & 1  & Sine\\
$\psi$     & $0$  &$ \pi $  & 1  & Uniform\\
$\phi$      & $0$  &$2 \pi $  & 1  & Uniform\\
$t_{\rm c}$          & \rm{trigger~time} - 0.1  & \rm{trigger~time} + 0.1  & \rm{s} & Uniform \\
$a_{\rm L}$          & $-0.005$  &$0.005$    & $c~\rm{s}^{-1}$ & Uniform\\
\toprule\end{tabular} \label{tab:table1}}
\end{center}
\end{table*}

\bibliographystyle{aasjournal}



\end{document}